\documentstyle[12pt]{article}
\begin{document}

\begin{center}

{\large\bf Rearrangement of the Fermi Surface of Dense Neutron
                          Matter \\
    and Direct Urca Cooling of Neutron Stars}

\vskip 0.5 cm
D.~N.~Voskresensky$^{1}$

{\small\it Moscow Institute for Physics and Engineering,
Moscow 115409, Russia}

\vskip 0.2 cm
V.~A.~Khodel$^{2}$ and M.~V.~Zverev

{\small\it Russian Research Center Kurchatov Institute,
Moscow 123182, Russia}

\vskip 0.2 cm
and

\vskip 0.2 cm
J.~W.~Clark

{\small\it McDonnell Center for the Space Sciences and
Department of Physics,
Washington University, \\ St. Louis, MO 63130 USA}

\vskip 0.5 cm
$^{1}${\small\it Gesellschaft f\"ur Schwerionenforschung
GSI, D-64220 Darmstadt, Germany}

$^{2}${\small\it McDonnell Center for the Space Sciences
and Department of
Physics, Washington University, \\ St. Louis, MO 63130 USA}

\end{center}

\vskip 0.5 cm

\begin{abstract}
It is proposed that a rearrangement of single-particle degrees of
freedom may occur in a portion of the quantum fluid interior of
a neutron
star.  Such a rearrangement is associated with the pronounced softening of
the spin-isospin collective mode which, under increasing density, leads
to pion condensation.  Arguments and estimates based on fundamental
relations of many-body theory show that one realization of this phenomenon
could produce very rapid cooling of the star via a direct nucleon Urca
process displaying a $T^5$ dependence on temperature.

\end{abstract}


\vskip 0.5 cm
\centerline{1.~INTRODUCTION}

\vskip 0.3 cm

The {\small EINSTEIN}, {\small EXOSAT}, and {\small ROSAT} orbiting
X-ray observatories have measured surface temperatures of certain neutron
stars and set upper limits on surface temperatures of others (\cite{Alpar}).
The data for the supernova remnants in 3C58, the Crab, and RCW103 indicate
relatively slow cooling, while that for Vela, PSR2334+61, PSR0656+14,
and Geminga point to substantially more rapid cooling.  In the so-called
standard scenario for neutron-star cooling, the primary role is played
by the modified Urca process
$(nn\rightarrow npe^-\bar{\nu}_e;\, npe^-\rightarrow nn\nu_e)$,
first considered by Bahcall \& Wolf (1965) and later reexamined by
Friman \& Maxwell (1979) in terms of an in-vacuum one-pion exchange
model.  Cooling simulations based on the results of these works play out
the slow scenario of thermal evolution and fail to explain the rapid
cooling of some stars.  This picture is profoundly altered when in-medium
effects are taken into account.  Modification of $NN$, $\pi N$, and
$K N$ interactions with increasing density may be so strong that pion
(\cite{Migdl,MSTV,Ph}) and kaon (\cite{Br94}) condensates form in the
interior region of a high-mass neutron star, leading to a dramatic
increase of the neutrino luminosity (\cite{MSTV,Max,Schaab}).

Here we shall demonstrate that softening of the spin-isospin (pion)
collective mode in dense neutron matter as predicted by Migdal (1978)
could give rise to a rearrangement of single-particle degrees
of freedom prior to the onset of pion condensation and open a new
channel of neutrino cooling of neutron stars, by giving access to
the {\it direct} Urca process.  This process does not involve a
neutron (or other) spectator and, if allowed, is an extremely
efficient cooling mechanism (\cite{Pethick}).

\vskip 0.5 cm
\centerline{2.~BUBBLE REARRANGEMENT OF THE FERMI SPHERE}

\vskip 0.3 cm
A rearrangement of single-particle degrees of freedom takes place if
the necessary condition for stability of the normal state of a Fermi
liquid is violated.  At $T=0$ this condition requires that the change
of the ground-state energy $E_0$ remain positive for any admissible
variation $\delta n(p)$ of the Landau quasiparticle distribution $n(p)$
away from the normal-state step-function distribution $\theta(p-p_F)$.
Formally,
\begin{equation}
\delta E_0 =\int \xi \left(p,n(p) \right) \delta n(p)
   {{\rm d}^3 p\over (2\pi)^3}>0 \, ,
\label{nsc}
\end{equation}
where $\xi(p,n(p))\equiv \varepsilon \left(p,n(p) \right)-\mu $ is the
energy of a quasiparticle relative to the chemical potential $\mu$.
The condition (\ref{nsc}) fails if a depression with $\xi<0$ forms
in the spectrum $\xi(p)$ at $p>p_F$; it likewise fails if there
arises an elevation with $\xi>0$ at $p<p_F$.   The rearrangement
is precipitated when the density $\rho$ reaches a critical value
$\rho_{cF}$ at which there emerges a bifurcation and a new root
$p=p_0$ of the relation
\begin{equation}
\xi\left( p,n(p);\rho_{cF}\right) =0 \, ,
\label{bif}
\end{equation}
which ordinarily serves merely to specify the Fermi momentum $p_F$.

The simplest kind of rearrangement of the momentum distribution $n(p)$
of quasiparticles of given spin projection retains the property
that its values are restricted to 0 and 1, but the Fermi sea
becomes doubly connected (\cite{deL}).  At densities
exceeding the critical value $\rho_{cF}$, the normal-state distribution
$\theta(p-p_F)$ is modified by the presence of a ``bubble,'' or absence
of particles, over the range $p_i < p < p_f < p_F$, with the inner
surface $p_i$ located relatively close to the origin and the Fermi
momentum $p_F$ readjusted to maintain the prescribed neutron density.
The distance $p_f-p_i$ between the two new Fermi surfaces lying
interior to $p_F$ can be estimated using the formula
$\xi(p\rightarrow p_0,n(p);\rho)=\xi_0(\rho-\rho_{cF}) -A(p-p_0)^2$,
where $A$ and $\xi_0$ are positive constants.  This formula embodies
the essential properties that $\xi(p)$ is negative
for any $p<p_F$ at $\rho < \rho_{cF}$ and that its maximum value first
reaches zero at $\rho=\rho_{cF}$.  Employing this parametrization
in the relation (\ref{bif}), it is found that no bifurcation point
exists for $\rho < \rho_{cF}$, whereas two solutions arise for
$\rho > \rho_{cF}$, with the distance between the two new Fermi
surfaces growing in proportion to $\sqrt{\rho-\rho_{cF}}$.

Were such a rearrangement to occur in the neutron subsystem of
neutron-star matter, the emergence of new Fermi surfaces situated at
lower momenta would permit the direct nucleon Urca process to operate
at a much lower density than hitherto considered possible
(\cite{Pethick,Lattimer}).  In this
process, the tandem reactions $n\rightarrow pe^-\bar{\nu}_e$ and
$pe^-\rightarrow n\nu_e$ are driven by thermal excitations.  The
condition of high degeneracy prevailing even in young neutron stars
implies that these excitations remain close to the Fermi surfaces
of the participants.  For the direct Urca mechanism to contribute
appreciably to neutron-star cooling, momentum conservation then
demands satisfaction of the triangle inequalities
$| p_{Fp}- p_{Fe}| \leq p_{Fn}\leq   p_{Fp}+   p_{Fe} $ among
the proton, electron, and neutron Fermi momenta.  With the conventional
singly-connected neutron Fermi sphere, these conditions are met only
at very high baryon densities where the proton Fermi momentum $k_{Fp}$
reaches sufficiently large values.  The best current estimates yield
a threshold baryon density of at least five times the saturation density
$\rho_0$ of symmetrical nuclear matter (\cite{Ph}).  On the other hand,
if the neutron subsystem undergoes a rearrangement that allows for
thermal excitations of neutron quasiparticles at much lower momenta
than $p_{Fn}$, specifically at $p_i$ and $p_f$ in the bubble
rearrangement scenario, the triangle inequalities are much more easily
satisfied and the direct Urca process is greatly facilitated.

We next provide a quantitative basis for this qualitative idea,
by appealing to established methods of microscopic many-body theory.
For a bifurcation point to arise in the solution of equation
(\ref{bif}), both the spectrum $\xi(p)$ and the scalar component
$f({\bf p}_1,{\bf p}_2;{\bf k}=0) $ of the Landau amplitude of the
quasiparticle interaction must depend strongly on momentum.  The
connection (\cite{Trio}, p.~37)
\begin{equation}
{\partial \xi_(p)\over \partial {\bf p}}=
{\partial \xi^0 (p)\over \partial {\bf p}}
+{1\over 2}{\rm Tr}_{\sigma}{\rm Tr}_{\sigma_1}
\int f({\bf p},\sigma,{\bf p}_1;\sigma_1, {\bf k}=0)
{\partial n(p_1)\over\partial {\bf p}_1}{{{\rm d}^3 p_1} \over{(2\pi)^3}}
\label{lnd}
\end{equation}
between these two quantities, wherein $\xi_0(p)$ is the free spectrum
and spin dependence is made explicit, is usually employed to find
the effective mass $M^*$ in terms of the first harmonic of the expansion
of the amplitude $f({\hat {\bf p}}_1 \cdot {\hat {\bf p}}_2;0)$
in Legendre polynomials.  We exploit this relation in the search for
new roots of equation (\ref{bif}).

Within the integral in (\ref{lnd}), the quasiparticle interaction
amplitude $f({\bf p}_1,{\bf p}_2;{\bf k}=0 )$ may be replaced by
$(M^*/M)\Gamma^k({\bf p}_1,{\bf p}_2 )$, the scattering amplitude
$\Gamma^k({\bf p}_1,{\bf p}_2)\equiv \Gamma({\bf p}_1,{\bf p}_2;{\bf k}=0 )$
modified by an effective-mass factor (\cite {Trio}).  The requisite
strong momentum dependences can arise if the system approaches
a second-order phase transition that occurs at some critical density
$\rho_c$ where a collective mode of frequency $\omega_s(k)$ collapses at wave
number $k=k_0$ and the corresponding susceptibility diverges along with
the scattering amplitude.  Near such a soft-mode critical point, the
singular part $\Gamma^s$ of the scattering amplitude
$\Gamma({\bf p}_1,{\bf p}_2; {\bf k})$ at momentum transfer ${\bf k}$
may be expressed quite generally by (\cite{dug})
\begin{equation}
{M^*\over M}\Gamma^s_{\alpha\kappa;\beta\lambda}({\bf p}_1,{\bf p}_2;
{\bf k}, \rho\rightarrow \rho_c)= -O_{\alpha\kappa}O_{\beta\lambda} D(k)+
O_{\alpha\lambda}O_{\beta\kappa}D(|{\bf p}_1-{\bf p}_2+ {\bf k}|)
\label{gams} \, ,
\end{equation}
where $O$ denotes the vertex determining the structure of the collective-mode
operator (e.g.\ $O=(\sigma\cdot{\bf n})\tau$ for the spin-isospin mode
with pion quantum numbers, where ${\bf n}$ is a unit vector along the
relevant momentum).  In deriving (\ref{gams}), antisymmetry of the
two-particle wave function under interchange of the coordinates and
spins of the two particles has been invoked.  The propagator $D$ may be
parametrized as $D(k)=\left[ \beta^2+\gamma^2(k^2/k^2_0-1)^2 \right]^{-1}$,
where $\beta(\rho)$ measures the proximity to the phase transition point,
with $\beta(\rho\to\rho_c)\sim \rho_c-\rho $ (cf.~Dyugaev (1976)).

The essential messages of the preceding development are that the
singular part
$ \Gamma^s({\bf p}_1,{\bf p}_2;{\bf k}=0)\sim D(|{\bf p}_1-{\bf p}_2|)$
of the scattering amplitude depends on the difference
$ {\bf p}_1-{\bf p}_2 $ and that as one approaches the soft-mode
phase transition point this dependence becomes quite strong.  We
assume that the remaining contributions to
$\Gamma({\bf p}_1,{\bf p}_2; {\bf k}=0)$ can be adequately
incorporated by renormalization of the chemical potential
$\mu$.  Equation (\ref{lnd}) is then easily integrated to produce an
explicit expression suitable for calculation of the single-particle
spectrum,
\begin{equation}
\xi(p)= \xi^0(p)
+{1\over 2} {\rm Tr} (O_{\alpha\lambda}O_{\lambda\alpha})
\int D(|{\bf p}-{\bf p}_1|) n(p_1){{{\rm d}^3 p_1}\over{(2\pi)^3}} \, .
\label{main}
\end{equation}

We now apply this equation to dense neutron matter in the vicinity of
the second-order phase transition associated with neutral pion
condensation, which is engendered by the softening of the spin-isospin
mode having $\pi^0$ quantum numbers (\cite {Migdl}).  It has been
predicted that the collapse of this mode will take place at a neutron
density $\rho_c=\rho_{c\pi}$ in the range (0.2 -- 0.5)~fm$^{-3}$
(roughly 1--3 times $\rho_0$), depending on theoretical assumptions
(\cite {MSTV,Ph}).  Unfortunately, there is as yet no
definitive microscopic treatment of neutron-star matter from which
one can extract or derive quantitatively reliable values for the
input parameters $\beta$, $\gamma$, and $k_0$ of our model.

In this situation, a reasonable strategy is to perform calculations based
on expression (\ref{main}) for several choices of the parameters of the
microscopic model.  Substituting (5) into relation (2), one finds the
critical density $\rho_{cF}$ for the onset of a bifurcation of the
latter equation.  For $\rho > \rho_{cF}$ this equation then determines
two new momenta $p_i$ and $p_f$ where $\xi(p)$ vanishes, which delimit
the bubble region of $n(p)$ and between which $\xi(p)$ is positive.
Representative numerical results for the spectrum $\xi(p)$ are plotted
in Fig.~1.  Results for the phase diagram of dense neutron matter
in the $\rho/\rho_0$ versus $\beta^2/m_\pi^2$ plane are displayed
in Fig.~2.  Different values of $\gamma$ are considered, while keeping
the parameter $k_0$ fixed at the value $0.9p_{Fn}$ suggested by
earlier numerical calculations (\cite{MSTV}).



It is evident that variation of the parameters $\beta$, $\gamma$,
and $k_0$ within sensible bounds can have strong effects on the phase
diagram and therefore on the extent of the phase with rearranged
quasiparticle occupation. Nevertheless, our numerical study has
documented three salient features of the bubble rearrangement.
{\it First}, the critical density $\rho_{cF}$ for the rearrangement is
less than the critical density $\rho_{c\pi}$ for pion condensation.
Since both phenomena stem from the strong momentum dependence of the
Landau amplitude $f({\bf p}_1,{\bf p}_2;{\bf k}\rightarrow 0)$,
rearrangement of the quasiparticle distribution may be regarded as
a {\it precursor} of pion condensation.  {\it Second}, the bifurcation
point corresponding to formation of a hole bubble in the neutron momentum
distribution is positioned at small momenta, $p_0< 0.2p_F$, irrespective
of the applicable value of $\rho_{c\pi}$.  {\it Third}, the spectrum
$\xi(p)$ shows a deep depression for $p \sim (0.5-0.6)p_F$.
And {\it fourth}, the ratios $\rho_{cF}/\rho_{c\pi}$ and
$p_0/p_F$ are insensitive to the actual value taken by $\rho_{c\pi}$
within the usual range of theoretical predictions.

Analogous considerations apply to the proton subsystem of neutron-star
matter, in which case one is dealing with the charged pion mode.
Estimates (\cite{Migdl,MSTV}) indicate that this mode is also
softened in dense matter, with a critical density not far from that
for neutral pion condensation.  One may then argue that, under the
influence of the strongly momentum-dependent external field provided
by the neutron medium, protons will leave the old Fermi sphere and
occupy states of relatively large momentum, $p \sim 0.5p_{Fn}$.  The
impact of this further rearrangement on the proton-neutron ratio
and on the rate of neutrino cooling requires a separate analysis,
which we defer.

Having laid the microscopic basis for a rearrangement of the neutron
Fermi surface that creates a bubble at low momenta in the Fermi sea, we
return to its most striking astrophysical implication.  Beyond the
bifurcation point, the triangle inequalities can now be satisfied without
the conventional requirement (\cite{Lattimer,Ph})) that the proton
fraction exceed some 11--14\%.  Accordingly, the direct Urca process
becomes active in the density regime just short of the threshold for
pion condensation.  At temperatures $T$ above the anticipated superfluid
phase transition, the $T$-dependence of the resulting neutrino emissivity
is determined through the usual expression
\begin{eqnarray}
\epsilon (n \rightarrow p e^- \bar{\nu})&=&{{2\pi}\over\hbar}
2G^2_F(1+3g^2_A)\sum_i\varepsilon_{\nu} n_n(1-n_p)(1-n_e) \nonumber \\
& & \quad \times \delta ({\bf p}_n-{\bf p}_p-{\bf p}_e-{\bf p}_{\nu})
  \delta(\varepsilon_n-\varepsilon_p-\varepsilon_e
-\varepsilon_\nu) \, ,
\end{eqnarray}
derived in quasiparticle approximation in terms of the occupations
$n_i$ and energies $\varepsilon_i$ of the reacting particles.  We
do not consider a renormalization of the coupling constants $G_F^2$
and $g_A^2$ due to medium effects, and the emissivity takes the
customary form (\cite{Lattimer})
\begin{equation}
\epsilon (n \rightarrow p e^- \bar{\nu})\simeq 1.2
\times 10^{27}{M^{\ast}_{n}\over M_n}{M^{\ast}_p\over M_p}
\left( \frac{\mu_e}{100 ~ \mbox{MeV}} \right)
T_{9}^6 \,\, \mbox{erg} \, \mbox{cm}^{-3} \, \mbox{sec}^{-1} \, ,
\label{emis}
\end{equation}
where $\mu_e$ is the electron chemical potential and the temperature
$T$ is measured in multiples of $10^9$ K.   Since the neutron and
proton effective masses $M^{\ast}_n$ and $M^{\ast}_p$ remain
$T$-independent, the bubble rearrangement serves only to turn
on the direct Urca process at a lower density, without altering
the familiar $T^6$ power-law behavior of the emissivity.

\vskip 0.5 cm
\centerline{3.~FERMION CONDENSATION}

\vskip 0.3 cm
Interestingly enough, there exists a more radical scenario for
rearrangement of the quasiparticle distribution, known as fermion
condensation (\cite{KS90,N92}).  In this case, the occupancy $n(p)$ may
be partial, i.e., it may lie between 0 and 1.  At $T=0$, the new
quasiparticle distribution $n(p)$ is to be found from the variational
condition
\begin{equation}
{\delta E_0 [n(p)] \over \delta n (p)}=\mu , \qquad p\in \Omega \, .
\label{var}
\end{equation}
The left-hand side of equation (\ref{var}) is just the quasiparticle
energy $\varepsilon(p)$.  Hence its coincidence with the chemical
potential $\mu$ means that the spectrum of single-particle excitations
will be dispersionless (i.e., the quasiparticle group velocity will
vanish) throughout the entire momentum domain $\Omega$ where
$\xi(p)=\varepsilon(p)-\mu=0$, and not merely at isolated points as
implied by equation (\ref{bif}).  The family of quasiparticles
having momenta $p \in \Omega$ is called the fermion condensate because of
a conspicuous analogy with the low-temperature Bose gas, in which the
energy of condensate particles is also equal to the chemical potential
$\mu$.  A key signature of fermion condensation has been observed
in strongly correlated electron systems (\cite{Nor}): flat portions
of the single-particle spectrum have been seen experimentally
in a number of high-temperature superconductors (\cite{Shen}).

The two types of rearrangement -- bubble formation and fermion
condensation -- can compete with each other in the density regime
just below the soft-mode phase transition.  Numerical studies (\cite{ZB})
demonstrate that fermion condensation wins the contest at nonzero $T$.
Let us suppose this is the case in the neutron-star medium, while
continuing to disregard nucleonic pairing phenomena.  Consistency of
the Fermi-Dirac form
$ n(p,T)=\left\{\exp \left[ \xi (p;n(p,T)) /T\right] +1\right\}^{-1}$ for
the quasiparticle distribution with the variational condition ({\ref{var})
requires that the spectrum $\xi(p,T)$ of the fermion-condensate
phase grows linearly with $T$ at low temperature, implying an effective
mass inversely proportional to $T$ (\cite{KS90,N92}).

Although the rearranged momentum distribution derived from equation
(\ref{var}) differs from the bubble configuration, its structure
will also admit thermal excitations at low neutron momenta.
Hence we may again expect most neutron stars to contain a region of
relatively moderate density, bounded below by $\rho_{cF}$ and above
by $\rho_{c\pi}$, in which the direct Urca process operates vigorously.
However, due to the new feature of a $T$-dependent neutron effective
mass, $M_n \propto 1/T$, we may anticipate an enhancement of the
neutrino emissivity relative to the standard result (\cite{Lattimer}),
corresponding to a $T^5$ rather than a $T^6$ dependence on the temperature.

\vskip 0.5 cm
\centerline{4.~CONCLUSIONS}

\vskip 0.3 cm
We have explored the possibility that an effective interaction
with strong momentum dependence gives rise to a rearrangement
of the neutron momentum distribution in neutron-star matter.  Two
plausible manifestations of this phase transformation -- creation of
a doubly-connected Fermi surface and fermion condensation -- have
been considered.  Both open the prospect that direct nucleon Urca
cooling is present in a density regime just below the threshold for
pion condensation and consequently at a density much lower
than previously estimated.  If a fermion condensate is formed, the
resulting neutrino emissivity is significantly larger than that
generated by the direct Urca process in normal matter.  Within
the affected density range, it would therefore dominate all
other proposed neutrino cooling mechanisms (\cite{Pethick}).  Future
studies along this line will focus on temperatures below the superfluid
transition and on the effect of the dramatically increased
emissivity on neutrino opacity.


\vskip 0.3 cm
This research was supported by NSF Grants PHY-9602127 and
PHY-9900713 (JWC and VAK) and by the McDonnell Center for the Space
Sciences (VAK).  We thank M.~Baldo, M.~Di Toro, and E.~E. Kolomeitsev
for fruitful discussions.  DNV expresses his appreciation for
hospitality and support provided by GSI Darmstadt.  MVZ
acknowledges the hospitality of INFN (Sezione di Catania).

\newpage

\centerline{FIGURE CAPTIONS}

\vskip 0.7 cm
Fig.~1. The dimensionless neutron spectrum
$y_n(p)=\xi_n(p)/(p_F^2/2M)$ at the
critical densities $\rho_{cF}$ corresponding to three different sets of
model parameters:
(a) $\gamma=1.25 m_\pi$, $k_0=0.9\,p_{Fn}$, $\beta^2=0.22\,m_{\pi}^2$
($\rho_{cF}\simeq 1.19\,\rho_0)$, (b) $\gamma=1.25 m_\pi$, $k_0=0.9\,p_{Fn}$,
$\beta^2=0.25\,m_{\pi}^2$ ($\rho_{cF}\simeq 1.76\,\rho_0)$,
(c) $\gamma=1.25 m_\pi$, $k_0=p_{Fn}$, $\beta^2=0.13\,m_{\pi}^2$
($\rho_{cF}\simeq 1.88\,\rho_0)$.  Two different positions of the
bifurcation point, namely $p_0=0$ (for parameter sets (a)
and (b)) and $p_0\simeq 0.12\,p_{Fn}$ (for set (c)), are indicated
by arrows.

\vskip 0.2 cm

Fig.~2. Phase diagram of neutron matter in the variables $\rho$
(measured in
$\rho_0$) and $\beta^2$ (measured in $m_{\pi}^2$), as calculated
for $k_0=0.9\,p_{Fn}$ and four different values of $\gamma$, which
(in $m_\pi$ units) label the corresponding the phase boundaries
separating the bubble phase (upper left) from the normal phase
(lower right).


\clearpage

\begin{thebibliography}{}
\bibitem{Trio}  Abrikosov, A. A., Gor'kov, L. P.,
\& Dzyaloshinski, I. E. 1965, Methods of Quantum Field Theory in Statistical
Physics (Englewood Cliffs: Prentice-Hall)
\bibitem{Ph}  Akmal, A., Pandharipande, V. R., \&
Ravenhall, D. G. 1998, Phys.~Rev. C, 58, 1804
\bibitem{Alpar}  Alpar, A., Kiziloglu, \"U., \& van
Paradijs, J. 1995, The Lives of the Neutron Stars (Dordrecht: Kluwer)
\bibitem{Bah}  Bahcall, J. N., \& Wolf, R. A.  1965,
Phys.~Rev.~B, 140, 1445
\bibitem[Brown 1994]{Br94}  Brown, G. E.  1994, Nucl.~Phys. A, 574, 217
\bibitem{deL} de Llano, M., \& Vary, J. P. 1979,
Phys.~Rev.~C, 19, 1083
\bibitem{dug}  Dyugaev, A. M.  1976, Sov. Phys. JETP, 43,
1247
\bibitem{FM}  Friman, B., \& Maxwell, O. V.  1979,
ApJ, 232, 541
\bibitem{KS90}  Khodel, V. A., \& Shaginyan,
V. R.  1990, JETP Lett., 51, 553
\bibitem{Lattimer}  Lattimer, J. M., Pethick,
C. J., Prakash, M., \& Hansel, P.  1991, Phys.~Rev.~Lett., 66, 2701
\bibitem{Max}  Maxwell, O. V., Brown, G. E.,
Campbell, D., Dashen, R., \& Manassah, J.  1977, ApJ, 216, 77
\bibitem{Migdl}  Migdal, A. B.  1978, Rev.~Mod.~Phys., 50, 107
\bibitem{MSTV}  Migdal, A. B., Saperstein, E. E.,
Troitsky, M. A., \& Voskresensky, D. N.  1990, Phys.~Rep., 192, 179
\bibitem{Nor} Norman, M. R.  1999, in High Temperature
Superconductivity, ed.~S. E. Barnes, J. Ashkenazi, J. L. Cohn,
\& F. Zuo (New York: AIP), 298
\bibitem{N92}  Nozi\`eres, P.  1992, J.~Phys.~I France, 2,
443
\bibitem{Pethick} Pethick, C. J.  1992, Rev. Mod. Phys., 64,
1133.
\bibitem{Schaab}  Schaab, Ch., Voskresensky, D. N.,
Sedrakyan, A. D., Weber, F., \& Weigel, M. K.  1997, A\&A, 321, 591
\bibitem{Shen}  Shen, Z. X., \& Dessau, D. S.  1995,
Phys.~Rep., 253, 1
\bibitem{Wam}  Wambach, J., Ainsworth, T. L., \&
Pines, D.  1993, Nucl.~Phys. A555, 128
\bibitem{ZB}  Zverev, M. V., \& Baldo, M.  1998,
JETP, 87, 1129

\end{thebibliography}
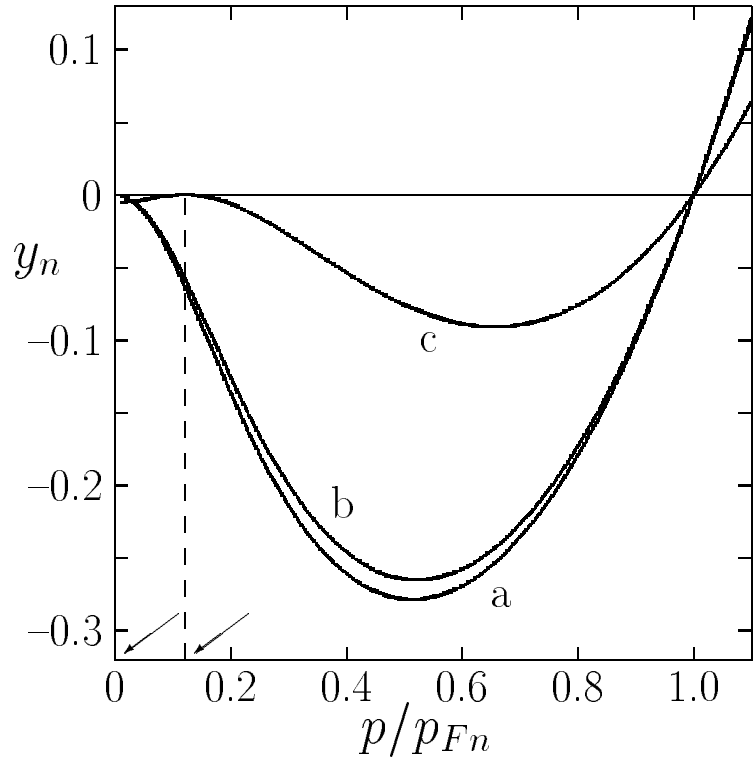
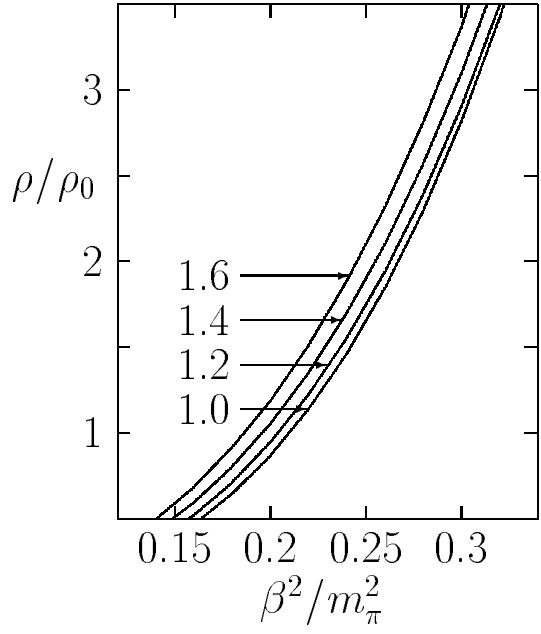
\end{document}